\documentclass[12pt, a4paper]{article}
\pdfoutput=1
\usepackage{graphicx}
\usepackage{amssymb}
\usepackage{amsmath}
\usepackage{bm}
\usepackage{color}
\usepackage{theorem}
\usepackage{subfigure}
\usepackage{colortbl}

\usepackage[sort&compress,numbers, merge]{natbib}

\definecolor{Orange}{cmyk}{0,0.61,0.87,0}
\definecolor{JungleGreen}{cmyk}{0.99,0,0.52,0}
\definecolor{OliveGreen}{cmyk}{0.64,0,0.95,0.40}
\definecolor{Brown}{cmyk}{0,0.70,1,0.40}
\definecolor{RoyalBlue}{cmyk}{0.71,0.53,0,0.12}
\definecolor{Gray}{cmyk}{0,0,0,0.40}
\definecolor{LightPink}{cmyk}{0.0,0.25,0,0}
\definecolor{LLightPink}{cmyk}{0.0,0.10,0,0}
\definecolor{LightBlue}{cmyk}{0.25,0,0,0}
\definecolor{LightGray}{cmyk}{0,0,0,0.2}

\newcommand{\Slash}[1]{{\ooalign{\hfil/\hfil\crcr$#1$}}}

\usepackage[colorlinks=true, linkcolor=black, citecolor=black,
urlcolor=black]{hyperref}

\begin{document}

\begin{titlepage}

\begin{flushright}
FTPI--MINN--16/03 \\
UMN--TH--3513/16
\end{flushright}

\vskip 1.35cm
\begin{center}

{\Large
{\bf
Testing the 2-TeV Resonance with Trileptons
}
}

\vskip 1.2cm

Arindam Das$^{a}$,
Natsumi Nagata$^{b}$,
Nobuchika Okada$^{a}$

\vskip 0.4cm

{\it $^a$Department of Physics and Astronomy, University of Alabama, \\
 Tuscaloosa, Alabama 35487, USA}\\[3pt]

{\it $^b$William I. Fine Theoretical Physics Institute, School of
 Physics and Astronomy, \\ University of Minnesota, Minneapolis,
 Minnesota 55455, USA} 

\date{\today}

\vskip 1.5cm

\begin{abstract}

 The CMS collaboration has reported a 2.8$\sigma$ excess in the search
 of the SU(2)$_R$ gauge bosons decaying through right-handed neutrinos into
 the two electron plus two jets ($eejj$) final states. This can be
 explained if the SU(2)$_R$ charged gauge bosons $W_R^\pm$ have a mass
 of around $2$~TeV and a right-handed neutrino with a mass of ${\cal
 O}(1)$~TeV mainly decays to electron. Indeed, recent results in several
 other experiments, especially that from the ATLAS diboson resonance
 search, also indicate signatures of such a $2$~TeV gauge 
 boson. However, a lack of the same-sign electron events in the CMS
 $eejj$ search challenges the interpretation of the right-handed
 neutrino as a Majorana fermion. Taking this situation into account, in
 this paper, we consider a possibility of explaining the CMS $eejj$
 excess based on the $\text{SU}(2)_L\otimes \text{SU}(2)_R\otimes
 \text{U}(1)_{B-L}$ gauge theory with pseudo-Dirac neutrinos. We find
 that both the CMS excess events and the ATLAS diboson anomaly can
 actually be explained in this framework 
 without conflicting with the current experimental bounds. This setup
 in general allows sizable left-right mixing in both the charged gauge
 boson and neutrino sectors, which enables us to probe this model
 through the trilepton plus missing-energy search at the LHC. It turns
 out that the number of events in this channel predicted in our model is
 in good agreement with that observed by the CMS collaboration. We also
 discuss prospects for testing this model at the LHC Run-II experiments.  

\end{abstract}

\end{center}
\end{titlepage}

\section{Introduction}
\label{sec:intro}

The CMS collaboration announced that they observed excess events in
their search for new massive charged gauge bosons $(W^\pm_R)$ associated
with the the SU(2)$_R$ gauge symmetry which decay into two leptons and
dijet through heavy right-handed neutrinos
\cite{Khachatryan:2014dka}. The excess was found in the invariant mass
distribution of the two electrons and dijet $(eejj)$ final states around
2~TeV, whose significance is 2.8$\sigma$. This signal, if confirmed,
certainly implies the presence of TeV-scale new physics. Various models
have been proposed so far to interpret this CMS excess; see,
\textit{e.g.}, Refs.~\cite{Bai:2014xba, Deppisch:2014qpa, Gluza:2015goa,
Dobrescu:2015qna, Dev:2015pga, Deppisch:2015cua, Awasthi:2015ota,
Dobrescu:2015jvn}. Among them, models based on the
$\text{SU}(2)_L\otimes \text{SU}(2)_R\otimes \text{U}(1)_{B-L}$ gauge
theory \cite{Pati:1974yy} are the simplest and most promising
candidates, since they contain right-handed neutrinos and $W^\pm_R$ as
their indispensable ingredients. Indeed, such models have attracted a
lot of attentions recently \cite{Dobrescu:2015qna, Dev:2015pga,
Deppisch:2015cua, Awasthi:2015ota, Dobrescu:2015jvn, Hisano:2015gna}
since they can explain possible anomalies observed in other (totally
independent) experiments, such as a 3.4$\sigma$ excess in the ATLAS
diboson resonance search \cite{Aad:2015owa}, an around $2\sigma$ excess
in the CMS dijet resonance search \cite{Khachatryan:2015sja}, and a
2.2$\sigma$ excess in the $W^\pm h$ channel where $W^\pm$ decays
leptonically and the Higgs boson $h$ decays into $b\overline{b}$
\cite{CMS:2015gla}. All of these results indicate the presence of
$W^\pm_R$ with a mass of around $2$~TeV.

If such a TeV-scale $W^\pm_R$ exists, in the $\text{SU}(2)_L\otimes
\text{SU}(2)_R\otimes \text{U}(1)_{B-L}$ models, we also expect that
there are  right-handed neutrinos whose masses are of ${\cal
O}(1)$~TeV. The presence of these right-handed neutrinos is desirable 
since we can exploit them to explain the CMS $eejj$ excess events. An
important caveat here is, however, that the CMS collaboration observed
only one same-sign electron event among all $14$ $eejj$ events
\cite{Khachatryan:2014dka}. This observation disfavors the conventional
$\text{SU}(2)_L\otimes\text{SU}(2)_R\otimes \text{U}(1)_{B-L}$ model
with an SU(2)$_R$ triplet Higgs field; in
this case, right-handed neutrinos are Majorana fermions, with which we
expect the same number of same-sign dilepton events as that of
the opposite-sign ones. In addition, TeV-scale right-handed Majorana
neutrinos are stringently restricted by the recent ATLAS
\cite{Aad:2015xaa} and CMS searches \cite{CMS:2015sea,
Khachatryan:2015gha} in the same-sign leptons plus dijet final
states. Therefore, it is required to extend this conventional model so
that it evades the above problems.

The inverse seesaw \cite{Mohapatra:1986aw} mechanism offers a promising
way to reconcile the difficulties. In this mechanism, three singlet
fermions are added to the neutrino sector on top of right-handed
neutrinos. Then, small lepton-number violation in the singlet mass terms
results in three light left-handed
neutrinos as well as heavy pseudo-Dirac neutrinos. Since a neutrino
which couples to $W^\pm_R$ is a pseudo-Dirac fermion, the lepton
number is approximately conserved in the process of $W^\pm_R$ decaying
to the neutrino, which accounts for a lack of same-sign
electron events in the CMS $eejj$ signals. Moreover, this mechanism has
an advantage in explaining small neutrino masses with TeV-scale
$\text{SU}(2)_L\otimes \text{SU}(2)_R\otimes \text{U}(1)_{B-L}$
symmetry. With such a low-scale symmetry-breaking of SU(2)$_R$, the
ordinary type-I seesaw mechanism \cite{Minkowski:1977sc} can yield
small neutrino masses only with very small Yukawa couplings
unless a specific mass structure is assumed
\cite{Pilaftsis:2003gt}, while the inverse seesaw mechanism allows the
couplings to be sizable. This feature is favorable when the model is
considered in the framework of grand unification \cite{Dev:2009aw} like
SO(10) models \cite{Georgi:1974my}.

In this paper, we consider an $\text{SU}(2)_L\otimes
\text{SU}(2)_R\otimes \text{U}(1)_{B-L}$ model that is extended to
accommodate the inverse seesaw mechanism. For recent work which
considers a similar model, see Ref.~\cite{Dev:2015pga}. It is found that
our model can actually realize the right number of $eejj$ signals
observed in the CMS experiment \cite{Khachatryan:2014dka}. A
characteristic feature of our model is that it allows sizable
left-right mixing in both the charged gauge boson and neutrino
sectors. Indeed, such a significant $W$--$W_R$ mixing is favored from the
viewpoint of the ATLAS diboson excess \cite{Aad:2015owa}. Moreover, the
inverse seesaw mechanism allows a large left-right neutrino mixing while
keeping neutrino masses tiny. 
In the presence of the left-right mixing, a heavy Dirac
neutrino can decay into not only the two leptons plus two jets final
states via a virtual $W_R$ exchange, but also into a lepton plus a
gauge/Higgs boson channels via the left-right mixing. Such decay
processes yield a trilepton plus missing energy signature, which is
regarded as the golden channel for probing heavy Dirac neutrinos at the
LHC \cite{delAguila:2008cj, Chen:2011hc, Das:2012ze, Das:2014jxa,
Das:2015toa}. We study the prediction of our model in this channel, and
find that the predicted number of events is in good agreement with the
result given by the CMS collaboration \cite{Chatrchyan:2014aea}. We
further discuss the future prospects for testing this model at the next
stage of the LHC run.

This paper is organized as follows. In the next section, we first
describe our model which we consider in this work. In
Sec.~\ref{sec:br}, we show the decay branching ratios of $W_R$ and
heavy Dirac neutrinos. Then, we study the collider signatures of our
model in Sec.~\ref{sec:LHC}. Section~\ref{sec:conclusion} is devoted to
conclusion and discussions.

\section{Model}

To begin with, we propose a model based on the $\text{SU}(3)_C \otimes
\text{SU}(2)_L\otimes \text{SU}(2)_R\otimes \text{U}(1)_{B-L}$ gauge
symmetry which has the structure of the inverse seesaw mechanism
\cite{Mohapatra:1986aw} in the neutrino sector. As in the Standard Model
(SM), left-handed quarks and leptons form $\text{SU}(2)_L$ doublet
fields:
\begin{equation}
 Q_{L_i} = 
\begin{pmatrix}
 u_{L_i} \\ d_{L_i}
\end{pmatrix}
~, ~~~~~~
 L_{L_i} = 
\begin{pmatrix}
 \nu_{L_i} \\ e_{L_i}
\end{pmatrix}
~,
\end{equation}
where $i = 1,2,3$ denotes the generation index. On the other hand,
right-handed fermions are embedded into the $\text{SU}(2)_R$ fundamental
representation as
\begin{equation}
 Q_{R_i} = 
\begin{pmatrix}
 u_{R_i} \\ d_{R_i}
\end{pmatrix}
~, ~~~~~~
 L_{R_i} = 
\begin{pmatrix}
 N_{R_i} \\ e_{R_i}
\end{pmatrix}
~.
\end{equation}
In addition, we introduce three gauge-singlet fermions $S_{L_i}$, which lead
to chiral partner fields of $N_{R_i}$ as we see below. 

The Higgs sector of this model contains two Higgs multiplets. One is an
$\text{SU}(2)_L \otimes \text{SU}(2)_R$
bi-doublet scalar field with zero $B-L$ charge, which breaks the
electroweak symmetry and thus plays a role of the SM Higgs field. We
denote it by $\Phi$ and its vacuum expectation value (VEV) by
\begin{equation}
 \langle \Phi \rangle =
\begin{pmatrix}
 v_u & 0 \\ 0 & v_d 
\end{pmatrix}
~,
\end{equation}
with
$v =\sqrt{v_u^2 + v_d^2} \simeq 174$~GeV. Moreover, to break the
SU(2)$_R$ symmetry, we introduce an SU(2)$_R$ doublet Higgs field $H_R$
with a $B-L$ charge $+1$, whose VEV is given by
\begin{equation}
\langle H_R\rangle =
 \begin{pmatrix}
  0 \\ v_R
 \end{pmatrix}
~.
\end{equation}
This breaks $\text{SU}(2)_L \otimes \text{SU}(2)_R \otimes
\text{U}(1)_{B-L}$ to $\text{SU}(2)_L \otimes \text{U}(1)_{Y}$.

With these particle contents, the interaction terms are generically given as
follows:
\begin{align}
 {\cal L}_{\text{int}} = &-y^Q_{ij} \overline {Q}_{R_i} \Phi Q_{L_j}
-\widetilde{y}^Q_{ij} \overline {Q}_{R_i} \widetilde{\Phi} Q_{L_j}
-y^L_{ij} \overline {L}_{R_i} \Phi L_{L_j}
-\widetilde{y}^L_{ij} \overline {L}_{R_i} \widetilde{\Phi} L_{L_j}
\nonumber \\
&- f_{ij} \overline{L}_{R_i} i\sigma_2 H_R^* S_{L_j} -\frac{1}{2}\mu_{ij}
 \overline{S_{L_i}^c} S_{L_j} +\text{h.c.}~,
\label{eq:lint}
\end{align}
where $\widetilde{\Phi} \equiv \sigma_2 \Phi^* \sigma_2$ with $\sigma_a$
($a=1,2,3$) being the Pauli matrices, and $c$ indicates the charge
conjugation. Note that the Majorana mass terms for right-handed
neutrinos $N_{R_i}$ are forbidden by the SU(2)$_R$ gauge symmetry. 
After the above Higgs fields develop the VEVs, these
interaction terms lead to the mass terms of the fermions. Here we
assume that these Yukawa couplings and the VEVs are appropriately
chosen so that the resultant mass terms agree to the observed quark and
lepton masses as well as the Cabibbo--Kobayashi--Maskawa (CKM) matrix
elements.\footnote{Note that the structure of the quark/lepton Yukawa couplings is the same
as that of the generic two-Higgs doublet model. Thus, we have more
degrees of freedom for the Yukawa couplings than those in,
\textit{e.g.}, the type-II two-Higgs doublet model. These extra degrees
of freedom are actually desirable since we can choose the Yukawa
couplings to account for the observed fermion masses and mixing even
though we take $v_u/v_d = {\cal O}(1)$; if we instead consider the type-II
two-Higgs doublet model like structure, then $v_u/v_d$ should be equal to
$m_t/m_b$ in order to explain the observed top-bottom mass ratio. 
} The mass matrix of the
neutrino sector is written as
\begin{equation}
 {\cal L}_{\text{mass}} = -\frac{1}{2} \overline{\psi^c_i} 
{\cal M}_{ij} \psi_j +\text{h.c.}~,
\end{equation}
where $\psi_i \equiv (\nu_{L_i}, N^c_{R_i}, S_{L_i})$, and 
\begin{equation}
{\cal M}_{ij} =
\begin{pmatrix}
 0 & M_D & 0 \\
 M_D^T & 0 & M_N^T \\
 0 & M_N & \mu
\end{pmatrix}
_{ij}
\equiv 
\begin{pmatrix}
 0 &{\cal M}_D \\ {\cal M}_D^T & {\cal M}_N
\end{pmatrix}
_{ij}
~,
\end{equation}
with 
\begin{align}
 \left(M_D^T\right)_{ij} &= y_{ij}^L v_u +\widetilde{y}_{ij}^L v_d~,
 \nonumber \\
 \left(M_N^T\right)_{ij} &= f_{ij}v_R ~.
\label{eq:massyukawanu}
\end{align}
Notice that the Majorana mass terms for $N_{R_i}$ are still not produced
due to the choice of the Higgs field that breaks the SU(2)$_R$
symmetry.\footnote{If we used an SU(2)$_R$ triplet Higgs field with two
unit of the $B-L$ charge to break the
SU(2)$_R$ symmetry, then we would generically obtain Majorana mass terms
for $N_{R_i}$.} 
Here, we assume a hierarchical structure among the mass parameters in
the matrix, \textit{i.e.}, $|\mu_{ij}| \ll |(M_D)_{ij}| \ll
|(M_N)_{ij}|$. The mass matrix ${\cal M}$ can be block diagonalized
by means of a unitary matrix. We obtain the mass matrix for light
neutrinos as  
\begin{equation}
 M_\nu \simeq -{\cal M}_D {\cal M}_N^{-1} {\cal M}_D^T 
\simeq M_D M_N^{-1} \mu (M_N^T)^{-1} M_D^T ~,
\end{equation}
while the other two classes of mass eigenvalues are given by $M_N\mp
\mu/2$. The latter can be regarded as pseudo-Dirac neutrinos for
$|\mu|\ll M_N$. Notice that small neutrino masses are guaranteed by the
smallness of $|\mu|$, and these masses vanish in the limit of $\mu\to
0$. In this limit, the theory recovers the lepton-number symmetry, which
results in three massless neutrinos and three heavy Dirac
neutrinos. Since the $\mu_{ij}$ term in Eq.~\eqref{eq:lint} does not
break any symmetry in our model, $\mu_{ij}$ in principle can have
arbitrary large value. We do not specify any mechanism to obtain a small
$\mu$ in this paper, though there have been several proposal to explain
the smallness of $\mu$ by exploiting spontaneous breaking of the
lepton-number symmetry \cite{GonzalezGarcia:1988rw}, extra dimensions
\cite{Park:2009cm}, or generation of $\mu$ through radiative corrections
\cite{Ma:2009gu}. Finally, we note in passing that an extremely small
$|\mu|$ allows the lepton Yukawa couplings $f_{ij}$ to be sizable, which
then indicates that the left-right mixing in the neutrino sector can
also be significant.

The VEV of $H_R$ gives masses to not only heavy neutrinos but also gauge
bosons associated with the broken symmetries. After the symmetry
breaking, we have massive charged and neutral gauge bosons, $W_R^\pm$ and
$Z_R$, whose masses are given by
\begin{align}
 m_{W_R} \simeq \frac{g_R}{\sqrt{2}} v_R ~, 
~~~~~~
 m_{Z_R} \simeq \frac{\sqrt{g_R^2+g^2_{B-L}}}{\sqrt{2}} v_R ~, 
\label{eq:wzprimass}
\end{align}
respectively. Here, the SU(2)$_R$ gauge coupling constant $g_R$ and the
$B-L$ gauge coupling constant $g_{B-L}$ are related to the U(1)$_Y$
gauge coupling constant $g^\prime$ by
\begin{equation}
 \frac{1}{g^{\prime 2}} = \frac{1}{g_R^2} + \frac{1}{g_{B-L}^2} ~,
\label{eq:gaugecouprel}
\end{equation}
which follows from
\begin{equation}
 Y = T_{R}^3 + \frac{B-L}{2} ~,
\end{equation}
with $Y$, $T_R^A$, and $B-L$ denote the hypercharge, the SU(2)$_R$
generators, and the $B-L$ charge, respectively. From the relation
\eqref{eq:gaugecouprel}, we find that there is a lower bound on the
value of $g_R$ to keep the $B-L$ coupling perturbative; for instance,
$g_{B-L} < 1$ ($4\pi$) leads to $g_R\gtrsim 0.39$ (0.36).

\begin{figure}[t]
\centering
\includegraphics[clip, width = 0.5 \textwidth]{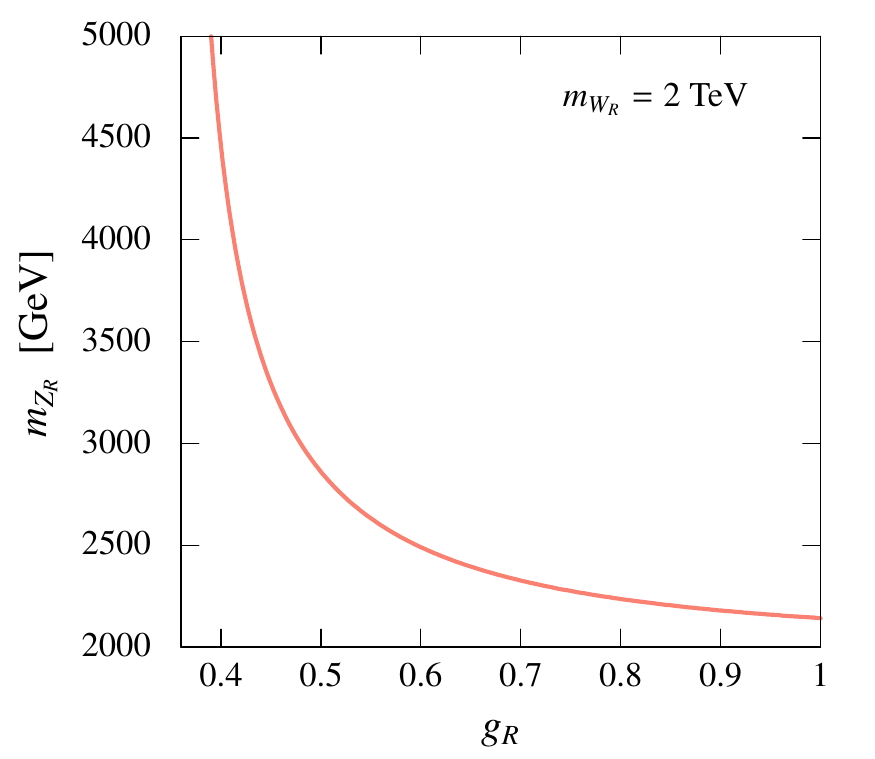}
\caption{Mass of $Z_R$, $m_{Z_R}$, as a function of the SU(2)$_R$ gauge
 coupling $g_R$. Here, we set $m_{W_R} =2$~TeV.} 
\label{fig:mzprime}
\end{figure}

As mentioned in Sec.~\ref{sec:intro}, recently there have been various
experimental observations which indicate the presence of $W_R^\pm$ with
a mass of around 2~TeV. Motivated by these observations, throughout this
paper, we assume $m_{W_R} \sim 2$~TeV. In this case, we can predict the mass of $Z_R$ as a
function of $g_R$ according to Eqs.~\eqref{eq:wzprimass} and
\eqref{eq:gaugecouprel}. In Fig.~\ref{fig:mzprime}, we plot $m_{Z_R}$ as
a function of $g_R$. Here, we set $m_{W_R} =2$~TeV. 
Currently, the most stringent limit on $Z_R$ is given by the ATLAS
collaboration using the $3.2$~fb$^{-1}$ data set at the center-of-mass
energy of $\sqrt{s} =13$~TeV \cite{ATLASdilepton} (see also the CMS
result \cite{CMS:2015nhc}).  
According to the ATLAS result, the production cross section of $Z_R$ times its
branching fraction into two leptons $\ell^\pm$ $(\ell = e,~\mu)$,
$\sigma (Z_R) \text{BR}(\ell^+ \ell^-)$ should be less than about 1~fb,
which gives a lower limit on the $Z_R$ mass of a several TeV. This limit
can easily be avoided if one takes $g_R \simeq 0.4$.

Since $m_{W_R} \sim 2$~TeV means $v_R ={\cal O}(1)$~TeV,
Eq.~\eqref{eq:massyukawanu} tells us that heavy pseudo-Dirac neutrinos
also have masses of ${\cal O}(1)$~TeV. To explain the CMS excess,
we take one of these heavy neutrinos to have a mass lighter than
$m_{W_R}$ and the others to have masses heavier than $m_{W_R}$ so that they
do not participate in the decay of $W_R$. We denote the former by $N_1$
and the latter by $N_2$ and $N_3$ in what follows. In addition, we
assume that $N_1$ mainly couples to electron; \textit{i.e.}, its
couplings with $\mu$ and $\tau$ leptons are negligible. In this setup,
${W_R}$ decays into a pair of right-handed quarks, $WZ$, $Wh$, or a
$N_1$ plus an electron. In the last case, the produced $N_1$
subsequently decays into an electron plus quarks via the
exchange of a virtual $W_R^\pm$. It can also decay into three leptons or
a lepton plus two quarks via
the $W^\pm$, $Z$, or the Higgs boson exchange if $N_1$ has a sizable
left-handed neutrino component or $W$--$W_R$ mixing is rather
large. Relevant formulae for the decay processes are summarized in
the subsequent section.

Finally, we give a brief discussion about the constraint on $W_R$ coming
from flavor physics. In this model, flavor-changing-neutral-current (FCNC)
processes can be induced by the exchange of $W_R$,\footnote{
As we discussed above, the structure of the Yukawa sector in our model
is similar to that in the generic two-Higgs-doublet model. Thus,
FCNC processes may also be induced by the exchange of the additional Higgs
bosons in general. In this paper, we simply assume that the Yukawa
couplings in our model are appropriately aligned so that FCNC processes
generated by the Higgs exchange are sufficiently suppressed. 
} which are severely
restricted from the low-energy precision flavor measurements. Among
them, the measurement of the $K_L$--$K_S$ mass difference
gives the most stringent bound on $m_{W_R}$, which is roughly
given by \cite{Zhang:2007da}
\begin{equation}
 m_{W_R} \gtrsim \left(\frac{g_{R}}{g_{L}}\right) \times
  2.5~\text{TeV} \simeq \left(\frac{g_R}{0.4}\right) \times 1.5~\text{TeV}
~.
\end{equation}
Hence, $W_R$ with a mass of around 2~TeV is still allowed by this bound
when we take $g_R\simeq 0.4$.

\section{Decay Branching Fractions}
\label{sec:br}

Here, we first summarize formulae relevant to the calculation of the partial
decay widths of $W_R^\pm$ and $N_1$. As mentioned above, ${W_R}$ decays
into a pair of right-handed quarks, $WZ$, $Wh$, or a $N_1$ plus an
electron. Among them, the $WZ$ and $Wh$ decay processes occur via the
mixing of $W_R$ with $W$ boson. Therefore, we begin with the discussion
on the $W$--$W_R$ mixing in our model. $W_R$ mixes with $W$ boson after
the bi-doublet Higgs field $\Phi$ acquires a VEV. The mass matrix of
these gauge bosons is given by
\begin{equation}
 {\cal L}_{\text{mass}} = (W^-_{L} ~W^-_{R})
\begin{pmatrix}
 \frac{g_L^2v^2}{2} & -\frac{g_L g_R v^2 \sin 2\beta}{2} \\
-\frac{g_L g_R v^2 \sin 2\beta}{2} & \frac{g_R^2}{2}\{v_R^2+v^2\}
\end{pmatrix}
\begin{pmatrix}
 W^+_{L} \\ W^+_{R}
\end{pmatrix}
~,
\end{equation}
where $W_L^\pm$ denote the SU(2)$_L$ gauge bosons, and $\tan\beta \equiv
v_d/v_u$. 
The mass matrix is diagonalized with an orthogonal matrix:
\begin{equation}
 \begin{pmatrix}
  W_{L}^+ \\ W_{R}^+ 
 \end{pmatrix}
=
\begin{pmatrix}
 \cos\phi_{LR}^W & -\sin\phi_{LR}^W \\ \sin\phi_{LR}^W & \cos \phi_{LR}^W
\end{pmatrix}
\begin{pmatrix}
 W_{1}^+ \\ W_{2}^+
\end{pmatrix}
~.
\end{equation}
Here, $W_{1}^+$ and $W_{2}^+$ are the mass eigenstates of the
charged gauge bosons. The corresponding eigenvalues are $m_W$ and
$m_{W_R}$, respectively, with $m_W \simeq g_L v/\sqrt{2}$ and $m_{W_R}$
given by Eq.~\eqref{eq:wzprimass}. In what follows, we refer to the
SU(2)$_L$-gauge-boson-like state $W_{1}^+$ as $W^+$. Since the mixing
angle $\phi_{LR}^W$ turns out to be extremely small in our scenario, we
denote $W^+_2$ also by $W_R^+$ unless otherwise noted. The mixing angle
$\phi_{LR}^W$ is then given by 
\begin{equation}
 \tan 2\phi_{LR}^W =\frac{2g_Lg_R v^2 \sin 2\beta}{g_R^2v_R^2-
(g_L^2-g_R^2)v^2}
\simeq 2\sin 2\beta
\biggl(\frac{g_R}{g_L}\biggr)\frac{m_W^2}{m_{W_R}^2} ~.
\end{equation}
 
The couplings of $W$ and $W_R$ to fermions are given as follows:
\begin{align}
 {\cal L}_{W_R ff} = &
\frac{g_L}{\sqrt{2}} \overline{u} \bigl(\cos \phi_{LR}^W \Slash{W}^+
-\sin\phi_{LR}^W \Slash{W}^{+}_R \bigr) P_L d 
+\frac{g_R}{\sqrt{2}} \overline{u} \bigl(\sin \phi_{LR}^W \Slash{W}^+
+\cos\phi_{LR}^W \Slash{W}^{ +}_R \bigr) P_R d \nonumber \\
 +&
\frac{g_L}{\sqrt{2}} \overline{\nu} \bigl(\cos \phi_{LR}^W \Slash{W}^+
-\sin\phi_{LR}^W \Slash{W}^{+}_R \bigr) P_L e
+\frac{g_R}{\sqrt{2}} \overline{N_1} \bigl(\sin \phi_{LR}^W \Slash{W}^+
+\cos\phi_{LR}^W \Slash{W}^{ +}_R \bigr) P_R e \nonumber \\
+&\text{h.c.} ~,
\end{align}
where we suppress the flavor indices for simplicity. 
In the mass eigenbasis, the $W_R$--$W$--$Z$ interaction is given by
\begin{align}
 {\cal L}_{W_R WZ} =& -ig_Z \sin \phi_{LR}^W\cos\phi_{LR}^W
(W_{\mu\nu}^+ W_R^{ -\mu} +W^{+}_{R\mu\nu}W^{-\mu}
-W^-_{\mu\nu}W_R^{+\mu}-W^{-}_{R\mu\nu}W^{+\mu})Z^\nu
\nonumber \\
&-ig_Z \sin\phi_{LR}^W\cos\phi_{LR}^W (W^+_\mu W^{-}_{R\nu} + W^{+}_{R\mu}
 W^-_\nu )Z^{\mu\nu} ~,
\label{eq:wrwz}
\end{align}
where $V_{\mu\nu} \equiv \partial_\mu V_\nu
- \partial_\nu V_\mu$ ($V = W$, $W_R$, or $Z$) and $g_Z\equiv
\sqrt{g^{\prime 2}+g_L^2}$. As for the $W_R W h$ coupling, we have
\begin{align}
 {\cal L}_{W_R Wh} = -\frac{1}{2\sqrt{2}}
[(g_L^2-g_R^2)\sin 2\phi_{LR}^W +2g_Lg_R \sin2\beta \cos 2\phi_{LR}^W]
vh (W^- W_R^{+}+W_R^{-}W^+) ~.
\end{align}

Now we evaluate the partial decay widths of $W_R$. For the fermion
channels, $W_R \to f \bar{f}^\prime$, we have
\begin{align}
 \Gamma (W_R^{+}\to u \bar{d})
&= \Gamma (W_R^{+}\to c \bar{s})
=\frac{g_R^2}{16\pi}
m_{W_R}~, \\
 \Gamma (W_R^{+}\to t \bar{b})
&=\frac{g_R^2}{16\pi}
m_{W_R} \left(1+\frac{m_t^2}{2 m_{W_R}^2}\right)
\left(1-\frac{m_t^2}{m_{W_R}^2}\right)^2~, \\
 \Gamma (W_R^{+}\to N_1 \bar{e})
&=\frac{g_R^2}{48\pi}
m_{W_R} \left(1+\frac{m_{N_1}^2}{2 m_{W_R}^2}\right)
\left(1-\frac{m_{N_1}^2}{m_{W_R}^2}\right)^2~,
\end{align}
where we have neglected the small mixing factor $\phi_{LR}^W$. 
For the $W_R\to WZ$ decay process, we have 
\begin{align}
 \Gamma(W_R^+ \to W^+ Z) 
=& \frac{g_R^2}{192\pi}\sin^2 (2\beta) m_{W_R}
\left(1-2
 \frac{m_W^2+m_Z^2}{m_{W_R}^2}+\frac{(m_W^2-m_Z^2)^2}{m_{W_R}^4}
\right)^{\frac{3}{2}}  \nonumber \\
&\times \left(
1+10 \frac{m_W^2+m_Z^2}{m_{W_R}^2} +
\frac{m_W^4 +10 m_W^2 m_Z^2 + m_Z^4}{m^4_{W_R}}
\right)~.
\end{align}
Here, notice that although the $W_R$--$W$--$Z$ coupling in
Eq.~\eqref{eq:wrwz} is suppressed by the small mixing angle
$\phi_{LR}^W$, the partial decay width of the $WZ$ channel does not
suffer from this suppression. This is because the high-energy behavior
of the longitudinal mode of $W_R$ gives an enhancement factor of $\sim
(m_{W_R}/m_W)^4$ and this compensates the suppression factor from the
mixing angle. Finally, the $W_R\to Wh$ decay width is given by
\begin{align}
 \Gamma(W_R^+ \to W^+ h) 
=& \frac{g_R^2}{192\pi}\sin^2 (2\beta) m_{W_R}
\left(1-2
 \frac{m_W^2+m_h^2}{m_{W_R}^2}+\frac{(m_W^2-m_h^2)^2}{m_{W_R}^4}
\right)^{\frac{1}{2}}  \nonumber \\
&\times \left(
1+\frac{10m_W^2-2m_h^2}{m_{W_R}^2} +
\frac{(m_W^2-m_h^2)^2}{m^4_{W_R}}
\right)~,
\end{align}
where we assume the decoupling limit for the Higgs bosons in our
model. Notice that in the large $m_{W_R}$ limit, 
\begin{equation}
  \Gamma(W_R^+ \to W^+ Z) \simeq 
 \Gamma(W_R^+ \to W^+ h)  ~,
\end{equation}
holds. This is a consequence of the equivalence theorem.

As seen above, the lightest Dirac neutrino $N_1$ is generated as a decay
product of $W_R$. The decay branching ratios of $N_1$ highly depend on
its mass and the left-right mixing in both the gauge boson and neutrino
sectors. When the mass of $N_1$ is rather large and the left-right
mixing is very small, the
three-body decay process via the virtual $W_R^+$ exchange is
dominant. The three-body decay width into an electron plus a pair of the
first/second generation quarks is given by \cite{Gluza:2015goa}
\begin{equation}
 \Gamma(N_1 \to \overline{q}q^\prime e^- ) = 
\frac{g_R^4}{2048 \pi^3} m_{N_1} F(x) ~,
\end{equation}
with $x = m_{N_1}^2/m_{W_R}^2$ and 
\begin{equation}
 F(x) = \frac{12}{x}\left[1-\frac{x}{2}-\frac{x^2}{6}+\frac{1-x}{x}
\ln (1-x)\right] ~.
\end{equation}
Here we neglect the quark and electron masses. For the $N_1 \to
\overline{b} t e^-$ decay channel, we have \cite{Dobrescu:2015jvn}
\begin{align}
 \Gamma (N_1 \to \overline{b} t e^-) = 
\frac{g_R^4}{2048 \pi^3} m_{N_1} F_t(x, y) ~,
\end{align}
where
\begin{align}
 F_t(x, y) &= \frac{12}{x} \biggl[
(1-y) -\frac{x}{2}(1-y^2) -\frac{x^2}{6}\left(
1-\frac{3}{2}y + \frac{3}{2}y^2 -y^3 
\right) \nonumber \\[3pt]
&-\frac{5x^3y}{8}(1-y^2) +\frac{x^4y^2(1-y)}{4}
-\frac{x^3y^2}{4}(4+x^2y)\ln y \nonumber \\
&+ \frac{1-x}{x} \ln \left(\frac{1-x}{1-xy}\right)
\left\{1-\frac{xy}{4}\left[4+x+x^2-x^3 y^2 (1+x)\right]\right\}
\biggr] ~,
\end{align}
with $y \equiv m_t^2/m_{N_1}^2$ ($m_t$ is the top mass). 
Of course, $F_t (x, y) \to F(x)$ as $y \to 0$. We note in
passing that the functions $F(x)$ and $F_t(x, y)$ also appear in the
calculation of the muon decay width \cite{Ferroglia:2013dga}.

On the other hand, if $m_{N_1}$ is relatively small and if
$\phi_{LR}^W$ or the mixing of $N_1$ with left-handed neutrinos $\nu_l$,
${\cal R}_{l1}$, is sizable, then the two-body decay processes become
dominant. In what follows, we assume that only the ${\cal R}_{e1}$
component can be sizable and the other flavor off-diagonal components,
${\cal R}_{\mu 1}$ and ${\cal R}_{\tau 1}$, are always negligible for
simplicity.\footnote{We here note that this assumption is consistent
with the experimental data of neutrino oscillations, as discussed in
Ref.~\cite{Das:2012ze}. } 
The relevant partial decay widths are then given as follows:
\begin{align}
 \Gamma (N_1 \to e^- W^+) &= \frac{g_L^2 |{\cal R}_{e1}|^2 + g_R^2 \sin^2
 \phi_{LR}^W}{64\pi} \frac{m_{N_1}^3}{m_W^2} 
\left(1-\frac{m_W^2}{m_{N_1}^2}\right)^2
\left(1+2 \frac{m_W^2}{m_{N_1}^2}\right)
~,\label{eq:n1lw}\\
\Gamma(N_1 \to \nu_e Z) &= \frac{g_Z^2 |{\cal R}_{e1}|^2}{128 \pi} 
\frac{m_{N_1}^3}{m_Z^2}\left(1-\frac{m_Z^2}{m_{N_1}^2}\right)^2 
\left(1+2 \frac{m_Z^2}{m_{N_1}^2}\right)
~,\\
\Gamma(N_1 \to \nu_e h) &= \frac{g_L^2 |{\cal R}_{e1}|^2}{128 \pi } 
\frac{m_{N_1}^3}{m_W^2} 
\left(1-\frac{m_h^2}{m_{N_1}^2}\right)^2 
~.
\end{align}

\begin{figure}[t]
\begin{center}
\subfigure[$\tan\beta$ dependence]
 {\includegraphics[clip, width = 0.48 \textwidth]{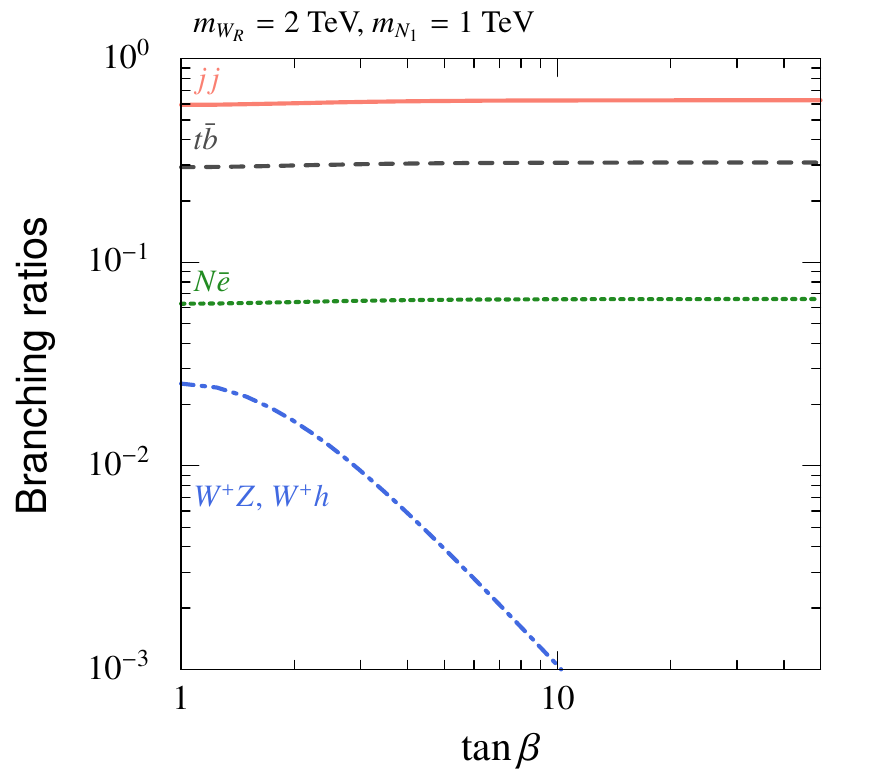}
 \label{fig:brtanb}}
\subfigure[$m_{N_1}$ dependence]
 {\includegraphics[clip, width = 0.48 \textwidth]{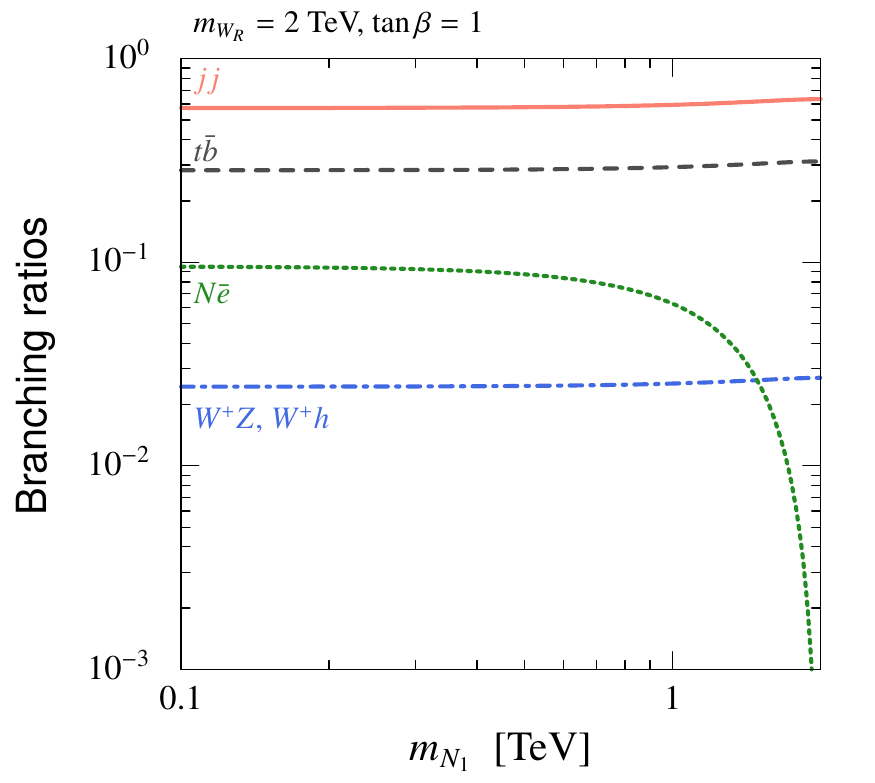}
 \label{fig:brmnr}}
\caption{Branching ratios of the $W_R^+$ decay as functions of $\tan
 \beta$ and $m_{N_1}$ in Figs.~\ref{fig:brtanb} and \ref{fig:brmnr},
 respectively. Here, we set $m_{W_R} = 2$~TeV. The red solid, black
 dashed, green dotted, and blue dash-dotted lines represent the
 branching fractions of the dijet, $t\bar{b}$, $N_1e^+$, and $W^+ Z$ and
 $W^+ h$ channels, respectively. $m_{N_1}$ is fixed to be 1~TeV in
 Fig.~\ref{fig:brtanb}, while $\tan \beta =1$ in Fig.~\ref{fig:brmnr}. }
\label{fig:brwr}
\end{center}
\end{figure}

By using the above formulae, we now evaluate the decay branching fractions
of $W_R$ and $N_1$. First, we show the branching ratios of the $W_R^+$
decay as functions of $\tan \beta$ and $m_{N_1}$ in
Figs.~\ref{fig:brtanb} and \ref{fig:brmnr}, respectively. Here, we set
$m_{W_R} = 2$~TeV. The red solid, black dashed, green dotted, and blue
dash-dotted lines represent the branching fractions of the dijet,
$t\bar{b}$, $N_1e^+$, and $W^+ Z$ and $W^+ h$ channels,
respectively. $m_{N_1}$ is fixed to be 1~TeV in Fig.~\ref{fig:brtanb},
while $\tan \beta =1$ in Fig.~\ref{fig:brmnr}. From these figures, we
find that about 10\% of $W_R$ decay into a pair of $N_1$ and
$e^+$ when $m_{N_1} \lesssim 1$~TeV. This decay branch hardly depends on
$\tan\beta$. Such a sizable decay fraction allows the model to explain the
CMS $eejj$ excess, as we will see below. The decay branch of $WZ$
channel, on the other hand, strongly depends on $\tan\beta$. In
particular, this model can explain the ATLAS diboson anomaly
\cite{Aad:2015owa} only if $\tan \beta$ is small; otherwise, the diboson
decay mode is almost negligible.

\begin{figure}[t]
\begin{center}
\subfigure[$\tan\beta = 1$]
 {\includegraphics[clip, width = 0.48 \textwidth]{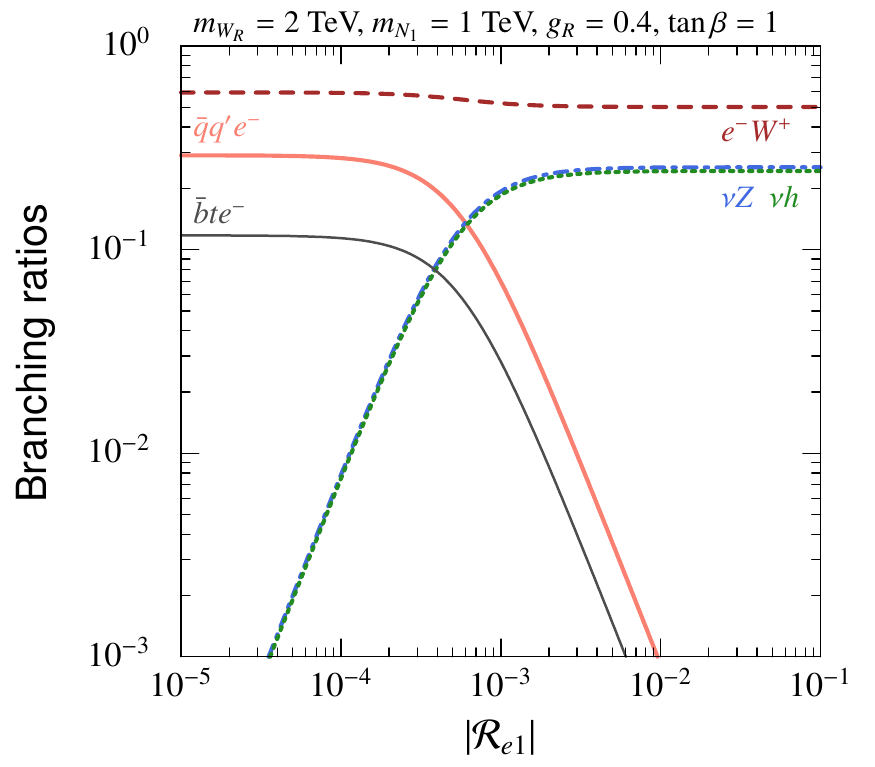}
 \label{fig:brnrre1tanb1}}
\subfigure[$\tan \beta =40$]
 {\includegraphics[clip, width = 0.48 \textwidth]{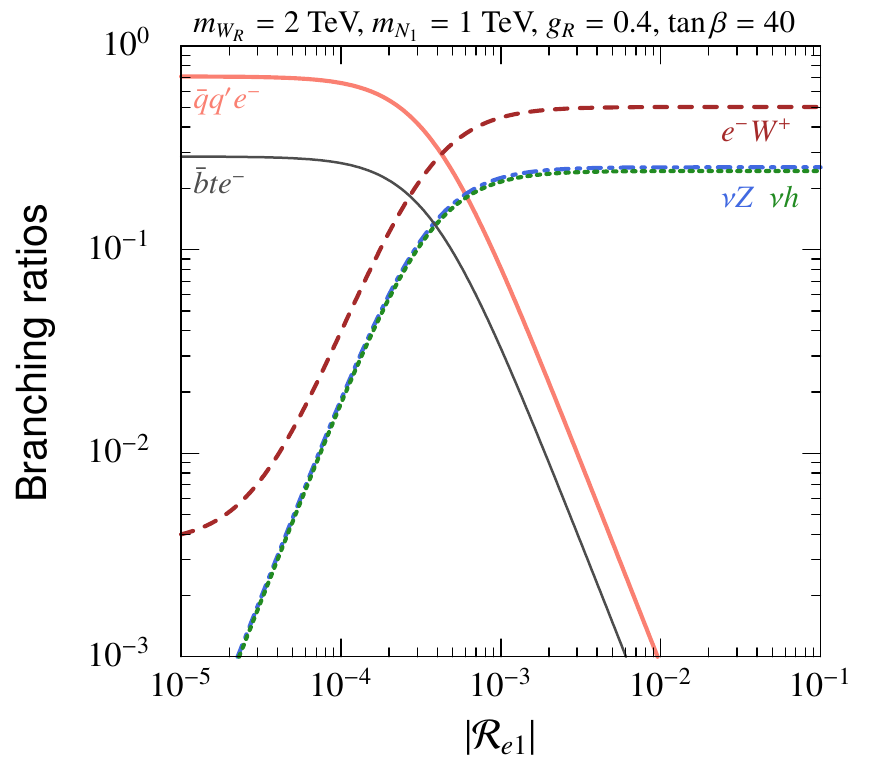}
 \label{fig:brnrre1tanb40}}
\caption{Branching ratios of the $N_1$ decay as functions of $|{\cal
 R}_{e1}|$. Here, we set $m_{W_R} = 2$~TeV, $m_{N_1}=1$~TeV, and $g_R = 0.4$. The red
 bold, black thin, brown dashed, green dotted, and blue dash-dotted
 lines represent the branching fractions of the $\bar{q}q^\prime e^-$,
 $\bar{b}te^-$, $e^- W^+$, $\nu Z$, and $\nu h$ channels,
 respectively. }
\label{fig:brnrre1}
\end{center}
\end{figure}

\begin{figure}[t]
\centering
\includegraphics[clip, width = 0.5 \textwidth]{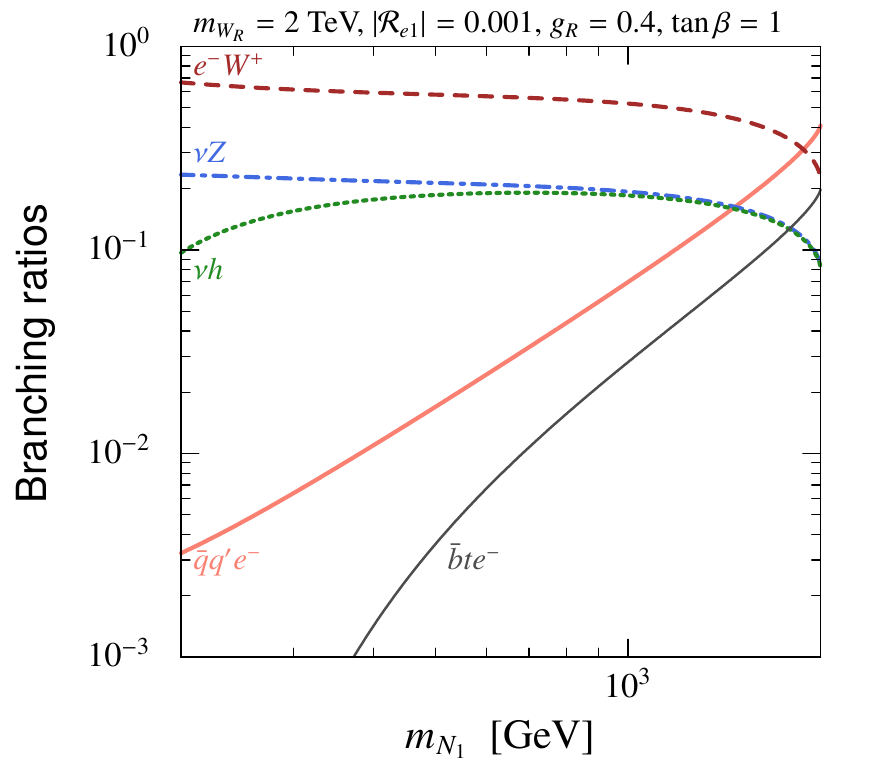}
\caption{Branching ratios of the $N_1$ decay as functions of
 $m_{N_1}$. Here, we set $m_{W_R} = 2$~TeV, $|{\cal R}_{e1}| = 0.001$,
 $g_R = 0.4$, and $\tan \beta =1$. The red
 bold, black thin, brown dashed, green dotted, and blue dash-dotted
 lines represent the branching fractions of the $\bar{q}q^\prime e^-$,
 $\bar{b}te^-$, $e^- W^+$, $\nu Z$, and $\nu h$ channels,
 respectively.} 
\label{fig:brnrmnr}
\end{figure}

Next, we evaluate the decay fractions of $N_1$. We plot the branching
ratios of the $N_1$ decay as functions of $|{\cal R}_{e1}|$ and
$m_{N_1}$ in Figs.~\ref{fig:brnrre1} and \ref{fig:brnrmnr},
respectively. Here, we set $m_{W_R} = 2$~TeV and $g_R = 0.4$. The red
bold, black thin, brown dashed, green dotted, and blue dash-dotted lines
represent the branching fractions of the $\bar{q}q^\prime e^-$,
$\bar{b}te^-$, $e^- W^+$, $\nu Z$, and $\nu h$ channels,
respectively. $m_{N_1}$ is fixed to be 1~TeV in Fig.~\ref{fig:brnrre1},
while $|{\cal R}_{e1}|=0.001$ in Fig.~\ref{fig:brnrmnr}. From
Fig.~\ref{fig:brnrre1}, we find that the three-body channels are sizable
only when $|{\cal R}_{e1}|$ is rather small. When $|{\cal R}_{e1}|$ is
large, the two-body decay channels become dominant as they are induced
via the left-right mixing in the neutrino sector in this case. However,
even in the small $|{\cal R}_{e1}|$ region, the branching fraction of
the $e^- W^+$ decay channel can still be sizable, depending on the value
of $\tan \beta$; this is because in 
this region the $e^- W^+$ decay is induced by the $W$--$W_R$ mixing.
As we see in
Sec.~\ref{sec:diboson}, $\tan \beta \simeq 1$ is favored in order to
explain the ATLAS diboson anomaly. In this case, the $e^- W^+$
channel is the dominant decay mode for any value of $|{\cal R}_{e1}|$,
as can be seen from Fig.~\ref{fig:brnrre1tanb1}. This allows us to test
our model with the trilepton plus missing energy channel.  On the other hand,
Fig.~\ref{fig:brnrmnr} shows that the branching fractions of the
three-body channels significantly depend on the mass of the right-handed
neutrino, while those of the two-body channels have relatively small
dependence on $m_{N_1}$.

\section{LHC Signatures}
\label{sec:LHC}

Now we study the LHC signature of our model. First, in
Sec.~\ref{sec:diboson}, we show the favored parameter space to explain
the excess events observed by the ATLAS collaboration in their diboson
resonance search \cite{Aad:2015owa}. Next, we consider 
the $eejj$ channel and determine the
parameters with which the model can explain the excess events observed
by the CMS collaboration \cite{Khachatryan:2014dka}. Then, in
Sec.~\ref{sec:trilepton}, we discuss prospects for probing our
model by using the trilepton plus missing energy searches.

\subsection{Diboson resonance search}
\label{sec:diboson}

The ATLAS collaboration has recently announced excessive events in the
diboson resonance search using fully hadronic decay channel
\cite{Aad:2015owa}. In this case, each gauge boson is reconstructed as a
fat jet since a gauge boson coming from a heavy resonance is highly
boosted so that the final-state two quarks from the gauge boson are
observed as a single large-radius jet. The ATLAS collaboration has observed a
narrow resonance around 2~TeV in the invariant mass distributions of two
fat jets, with its local significance of 3.4~$\sigma$ in the $WZ$
channel. The CMS collaboration also found a small excess around 1.9~TeV
\cite{Khachatryan:2014hpa} in a similar analysis. Recently, the ATLAS
collaboration \cite{Aad:2015ipg} combined the results of
searches for diboson resonances decaying into leptonic \cite{Aad:2014pha},
semi-leptonic \cite{Aad:2014xka, Aad:2015ufa}, and hadronic final states
\cite{Aad:2015owa}, and still found a 2.5$\sigma$ deviation from the SM
prediction. Taking into account these results, as well as those from the
CMS semileptonic search \cite{Khachatryan:2014gha}, the authors in
Ref.~\cite{Dias:2015mhm} have found that the above results are well
fitted with a 2~TeV $W_R$ whose production
cross section, $\sigma (pp\to W_R)$, times the branching fraction of the
$WZ$ decay channel, $\text{BR}(W_R \to WZ)$, is
\begin{equation}
 \sigma (pp\to W_R) \times \text{BR}(W_R \to WZ) = 4.3^{+2.1}_{-1.5} ~
  \text{fb} ~.
\label{eq:wzobs}
\end{equation}
We further note that the 13~TeV diboson resonance searches from both the
ATLAS \cite{ATLAS13TeVdiboson} and CMS \cite{CMS:2015nmz} collaborations
are found to be still too weak to constrain these possible anomalies
observed at the LHC Run--I.

Let us see if our model can reproduce the required value of $ \sigma
(pp\to W_R) \times \text{BR}(W_R \to WZ)$ given in
Eq.~\eqref{eq:wzobs}. We compute the production cross section of a 2~TeV
$W_R$ at
$\sqrt{s} = 8$~TeV by using {\tt MadGraph5} \cite{Alwall:2014hca} as
\begin{equation}
 \sigma (pp\to W_R) \simeq 90\times \left(\frac{g_R}{0.4}\right)^2
  \text{fb} ~.
\end{equation} 
Here, we re-scale the cross section by the so-called $k$ factor, $k
\simeq 1.3$ \cite{Cao:2012ng, Jezo:2014wra}, to include the effects of
the higher-order QCD corrections. To obtain the value in
Eq.~\eqref{eq:wzobs}, therefore, we need
\begin{equation}
 \text{BR}(W_R \to WZ) = 4.8^{+2.3}_{-1.7}\times 10^{-2}~,
\end{equation}
for $m_{W_R} = 2$~TeV and $g_R = 0.4$. From Fig.~\ref{fig:brwr}, we find
that this model can explain a part of the diboson excess only if $\tan\beta
\simeq 1$. This observation motivates us to consider the $\tan\beta
\simeq 1$ case. In this case, the left-right mixing in the gauge boson
sector is sizable, which plays an important role in the phenomenology of
the $N_1$ decay as we have seen in the previous section. 

Although our setup discussed here predicts a smaller number of events in
the diboson channel than the observed one, our model still may explain
all of the events with the $W_R$. For instance, by enhancing the
production cross section of $W_R$, we may increase the number of
events. This can be realized if we consider a slightly lighter $W_R$
(note that we cannot enhance the production cross section by using a
larger value of $g_R$ as it predicts a too light $Z_R$, as can be seen
from Fig.~\ref{fig:mzprime}); for example, we obtain $\sigma (pp\to W_R)
\simeq 130$~fb for $m_{W_R} = 1.9$~TeV and $g_R = 0.4$. On the other
hand, for a 1.9~TeV $W_R$, $\sigma (pp\to W_R) \times \text{BR}(W_R \to
WZ) = 5.3^{+2.3}_{-2.0} ~\text{fb}$ is favored from the experiments
according to Ref.~\cite{Dias:2015mhm}. This means $\text{BR}(W_R \to WZ)
= 4.1^{+1.9}_{-1.5}\times 10^{-2}$, which is relatively close to the model
prediction for $\tan\beta = 1$. Another way is to introduce an extra
Higgs field, \textit{e.g.}, an SU(2)$_R$ triplet Higgs field, which
gives an additional contribution to the $Z_R$ mass. In this case, we may
take a larger value of $g_R$ with keeping $m_{Z_R}$ large enough. By
taking the couplings of the additional Higgs field with the fermions in
our model (especially with right-handed neutrinos) sufficiently small,
we can keep heavy neutrinos pseudo-Dirac. Anyway, given the small
statistics at present, it is unclear whether our model can explain the
diboson anomaly without going beyond the minimal setup or not. This
situation should be settled by the LHC Run--II experiments in the near future.

There are several other decay channels which may constrain a 2~TeV
$W_R$. Figure~\ref{fig:brwr} shows that $W_R$ mainly decays into light
quarks, and thus dijet resonance searches can give a strong limit on the
production of $W_R$. At present, the ATLAS dijet resonance search
based on the $3.6$~fb$^{-1}$ data at the 13~TeV run gives the severest
limit \cite{ATLAS:2015nsi}: $\sigma (pp\to W_R) \times {\cal A}\times
\text{BR} (W_R\to jj) \lesssim 180$~fb with ${\cal A} \simeq 0.4$ being the
acceptance. The CMS limit is less severe than the ATLAS one because of
the smaller number of integrated luminosity
\cite{Khachatryan:2015dcf}. On the other hand, the production cross
section of a $W_R$ at $\sqrt{s}=13$~TeV is evaluated as $\sigma (pp\to
W_R) \simeq 557$~fb for $m_{W_R} = 2$~TeV and $g_R = 0.4$. Here, we have
used the $k$-factor of $k=1.2$ \cite{Cao:2012ng, Jezo:2014wra}. Hence,
the present ATLAS bound \cite{ATLAS:2015nsi} reads $\text{BR} (W_R\to
jj) \lesssim 0.81$, which is satisfied in our model as can be seen from
Fig.~\ref{fig:brwr}. The third-generation-quark resonance search can
also restrict this model. The strongest limit is currently given by the
CMS collaboration based on the 8~TeV run \cite{Khachatryan:2015edz}:  
$\sigma (pp\to W_R) \times \text{BR} (W_R\to tb) \lesssim 40$~fb for a
2~TeV $W_R$, which leads to $\text{BR} (W_R\to tb) \lesssim 0.44$ for
$g_R = 0.4$. Our model prediction is $\text{BR} (W_R\to tb) \simeq 0.3$,
which is below the present limit.

Finally, we comment on the indirect limit on the $W$--$W_R$ mixing from
the electroweak precision measurements. As seen above, to explain the
ATLAS diboson anomaly in our model, $\tan\beta \simeq 1$ is required,
which implies that the $W$--$W_R$ mixing angle should be ${\cal
O}(10^{-3})$. This size of the $W$--$W_R$ mixing potentially
conflicts with the electroweak precision measurements. Here, note that
we cannot use the $S$ and $T$ parameters \cite{Peskin:1990zt} to assess
the consistency of our model with the electroweak precision
measurements, 
since our model also contains a $Z_R$ and it modifies
the $Z$-boson coupling to the SM fermions at tree level through the
$Z$--$Z_R$ mixing. Instead, we need to carry out a complete parameter
fitting onto the electroweak observables. Such a parameter fitting is
done in Refs.~\cite{Cao:2012ng, Hsieh:2010zr} and it is found
that a 2~TeV $W_R$ with an ${\cal O}(10^{-3})$ $W$--$W_R$ mixing is
actually consistent with the electroweak precision experiments.

\subsection{$eejj$ Channel}
\label{sec:eejj}

Next, we discuss the $eejj$ channel. The CMS collaboration has observed
a 2.8$\sigma$ anomaly in this channel \cite{Khachatryan:2014dka} with
the 19.7~fb$^{-1}$ 8~TeV data, which also indicates the presence of
$W_R$ with a mass of around 2~TeV. 14 events are observed around $2$~TeV, while 4 events are
expected from the SM backgrounds. Among the 14 events, only one event
consists of same-sign dielectron, while the rest of 13 events
include opposite-sign electrons. The number of the same-sign dielectron
events due to the SM backgrounds is expected to be ${\cal O}(0.5)$;
thus, this observation is totally consistent with a hypothesis that all
of the signal events consist of opposite-sign dielectron events.
The signal acceptance ${\cal A}$ is
listed in Ref.~\cite{Khachatryan:2014dka}; for instance, for $m_{W_R} =
2$~TeV and $m_{N_1} = 1$~TeV, we have ${\cal A} = 0.784\pm 0.009$. This
implies that if the signal cross section of the $eejj$ channel is $\simeq
0.65$~fb, then the predicted number of events falls
right in the middle of the observed number. 

In our model, the $eejj$ decay process is induced via the virtual
$W_R$ exchange by a $N_1$, 
\begin{equation}
 W_R \to e N_1 \to e e W_R^* \to eejj ~,
\end{equation}
as well as via the on-shell $W$ which is a decay product of $N_1$:
\begin{equation}
  W_R \to e N_1 \to e e W \to eejj ~.
\end{equation}
Notice that we expect opposite-sign electrons in the final state, rather
than same-sign dielectron, since lepton-number violation is
significantly suppressed by the very small mass parameters $\mu_{ij}$ in
our model. This is consistent with the CMS observation.

\begin{figure}[t]
\begin{center}
\subfigure[$\tan\beta = 1$]
 {\includegraphics[clip, width = 0.48 \textwidth]{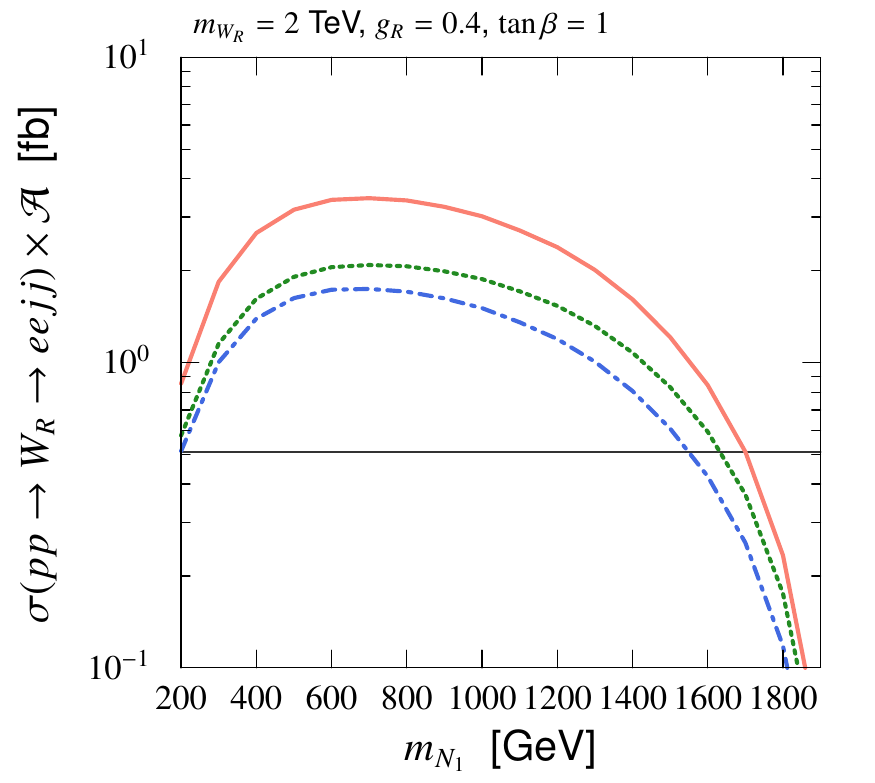}
 \label{fig:eejj1}}
\subfigure[$\tan \beta =40$]
 {\includegraphics[clip, width = 0.48 \textwidth]{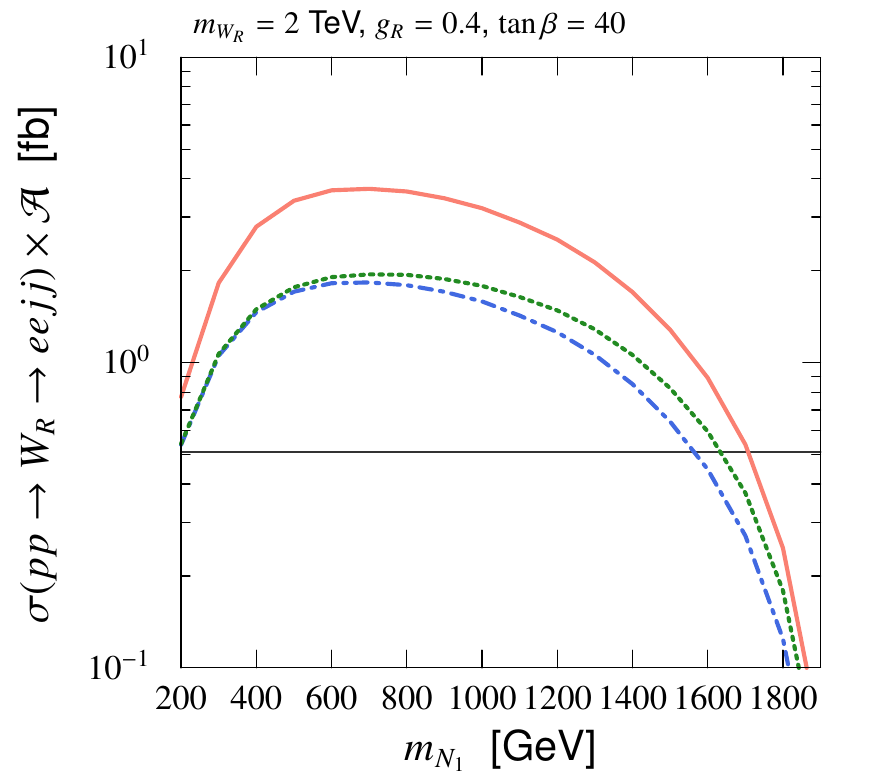}
 \label{fig:eejj40}}
\caption{Signal cross section for the $eejj$ channel times the
 acceptance ${\cal A}$ as functions of $m_{N_1}$. Here, we set $m_{W_R}
 = 2$~TeV and $g_R = 0.4$. The red solid, green dotted, and blue
 dash-dotted lines show the cases of $|{\cal R}_{e1}| = 10^{-4}$, $10^{-3}$,
 and $10^{-2}$, respectively. The horizontal gray line corresponds to 10
 events for an integrated luminosity of 19.7~fb$^{-1}$. }
\label{fig:eejj}
\end{center}
\end{figure}

In Fig.~\ref{fig:eejj}, we plot the signal cross section for the $eejj$
channel times the acceptance ${\cal A}$ as functions of
$m_{N_1}$. Here, we set $m_{W_R} = 2$~TeV and $g_R = 0.4$. The red
solid, green dotted, and blue dash-dotted lines show the cases of $|{\cal
R}_{e1}| = 10^{-4}$, $10^{-3}$, and $10^{-2}$, respectively. The
acceptance is taken from Ref.~\cite{Khachatryan:2014dka}. From these
plots, we find that although the decay branching ratios of $N_1$
significantly depend on $\tan \beta$ as shown in Fig.~\ref{fig:brnrre1},
the signal cross section for the $eejj$ mode does not depend on $\tan
\beta$ so much; if one takes $\tan \beta$ large, the $eW$ decay channel
of $N_1$ could be subdominant, but in this case the three-body $ejj$
decay mode becomes dominant, which makes the total signal cross
section for the $eejj$ decay channel almost unchanged. We also show the
value of the signal cross section which corresponds to 10 events for an
integrated luminosity of 19.7~fb$^{-1}$ by the horizontal gray line in
this figure. It is found that the observed event number is reproduced if
$m_{N_1}$ is in the range of $\sim 1.5$--1.7~TeV. We however note that
because of the low statistics we expect a large uncertainty in the
extraction of the favored signal cross section. Furthermore, our
computation also suffers from uncertainty resulting from the estimation
of the acceptance. In our analysis, we took the acceptance rate given by
the CMS collaboration \cite{Khachatryan:2014dka}. However, this
acceptance is estimated for the three-body decay of $N_1$ via
the off-shell $W_R$ exchange process. On the other hand, in our model,
the two-body decay of $N_1$ into $eW$ also gives rise to the $eejj$
final state. This contribution may result in a different value of
acceptance ${\cal A}$. Considering these possible uncertainties, we
conclude that at present any values of $m_{N_1} \sim 1$~TeV may be
consistent with the CMS $eejj$ search result.

\subsection{Trilepton Channel}
\label{sec:trilepton}

Now let us discuss possibilities to probe our model in the trilepton
plus large missing energy mode. As we have seen in
Sec.~\ref{sec:diboson}, $\tan\beta \simeq 1$ is favored in order to
explain the ATLAS diboson anomaly. In this case, the dominant decay mode
of $N_1$ is always the $eW$ final state. This state can subsequently decay into
the three charged leptons plus a light neutrino final state. Therefore,
our setup discussed so far in general predicts a sizable signal rate in
the trilepton plus large missing energy searches. 

To illustrate this, we compare the prediction of our model with the
CMS result of the search for the trilepton plus missing energy signatures at
the center-of-mass energy of $\sqrt{s}=8$~TeV with the 19.5~fb$^{-1}$
integrated luminosity \cite{Chatrchyan:2014aea}. As we have assumed
above, the flavor-violating processes are negligible in our
setup. Hence, we focus on events which contain an opposite-sign
same-flavor (OSSF) lepton pair. This category is called OSSF1 in
Ref.~\cite{Chatrchyan:2014aea}. Moreover, since $N_1$ only couples to an
electron, this pair should be $e^+ e^-$.\footnote{As we will state soon
below, we veto events if the invariant mass of any pair of OSSF charged
leptons is reconstructed to be around the $Z$-boson 
mass. This rejects the $N_1 \to \nu Z$ events, and thus we do not expect
final states which include a pair of $\mu^+ \mu^-$. } Therefore, the
trilepton events we consider below include either $e^+ e^- e^\pm$ or $e^+ e^-
\mu^\pm$.

In our analysis, we generate the trilepton plus missing energy events
using {\tt MadGraph5} \cite{Alwall:2014hca} and evaluate the parton-level
cross sections with the {\tt CTEQ6L} parton distribution function set
\cite{Pumplin:2002vw}. The cross sections are multiplied by the
$k$-factor of $k =1.3$ \cite{Cao:2012ng, Jezo:2014wra}. 
The showering and hadronization are executed with
{\tt PYTHIA6.4} \cite{Sjostrand:2006za}, while we use {\tt DELPHES3}
\cite{Ovyn:2009tx} for the detector simulation. Jet-clustering is
performed with {\tt FastJet2} \cite{Cacciari:2008gp} based on the
anti-$k_{\rm T}$ algorithm with a distance parameter of $0.5$. We impose the
same criterion for the event selection as those used in
Ref.~\cite{Chatrchyan:2014aea}: 
\begin{itemize}
 \item Electrons and muons are required to satisfy that their transverse momentum
       $p_{\rm T}$ be larger than 10~GeV and the magnitude of their
       pseudo-rapidity $\eta$ be smaller than 2.4. They should be
       separated from each other by $\Delta R \equiv \sqrt{(\Delta
       \eta)^2 + (\Delta \phi)^2} > 0.1$, where $\phi$ is the azimuthal
       angle. 

 \item At least one electron or muon should have $p_{\rm T} > 20$~GeV. 

 \item Jets should satisfy $p_{\rm T} > 30$~GeV and $|\eta| < 2.5$.
       They are required to be separated from a lepton by $\Delta R >
       0.3$. 

 \item For each event, we construct OSSF charged
       leptons $\ell^+ \ell^-$ $(\ell = e, \mu)$ and require that the
       invariant mass of these charged leptons, $m_{\ell^+ \ell^-}$,
       should be $\geq 12$~GeV. 

 \item We reject the ``on-Z'' events in which a pair of OSSF charged
       leptons yields $75 < m_{\ell^+ \ell^-} <105$~GeV.  

\end{itemize}
Then, we classify each event into several categories according to
Ref.~\cite{Chatrchyan:2014aea}. Firstly, we divide all events into two
classes: one with the scalar sum of jet transverse momentum, $H_{\rm
T}$, being $H_{\rm T} > 200$~GeV and the other with $H_{\rm T} <
200$~GeV. Secondly, we divide each class in terms of the missing
transverse energy $E_{\rm T}^{\rm miss}$: $E^{\rm miss}_{\rm T} >
100$~GeV, $50<E^{\rm miss}_{\rm T}<100$~GeV, or $E_{\rm T}^{\rm miss} <
50$~GeV. Here, $E_{\rm T}^{\rm miss}$ is the magnitude of the vector
sum of the transverse momenta. Finally, if all possible OSSF pairs
give $m_{\ell^+ \ell^-} > 105$~GeV ($m_{\ell^+ \ell^-} < 75$~GeV), then
the corresponding event is called an above-Z (below-Z) event.

\begin{table}[t!]
 \begin{center}
\caption{Simulated number of events in our model for the 8~TeV run with
  an integrated luminosity of 19.5~fb$^{-1}$. Here, we set $m_{W_R} =
  2$~TeV, $g_R =0.4$, ${\cal R}_{e1} = 10^{-3}$,
  and $\tan \beta =1$. }
\label{tab:simulation}
\vspace{5pt}
\begin{tabular}{ll|cccc}
\hline
\hline
Category & $m_{\ell^+ \ell^-}$& $m_{N_1} = 1$~TeV & 1.6~TeV &
Observed& Expected \\ \hline
\rowcolor{LightGray}
$H_{\rm T} > 200$~GeV
&& &  && \\
\hline
$E_{\rm T}^{\rm miss} > 100$~GeV &Above-Z &1.86 & 0.85&5 &$3.6\pm 1.2$ \\
&Below-Z &0 & 0& 7& $9.7\pm 3.3$ \\
\hline
$50<E_{\rm T}^{\rm miss} < 100$~GeV & Above-Z &0.22 &0.02& 4& $5.0 \pm 1.6$ \\
&Below-Z &0&0& 10& $11.0\pm 3.8$ \\
\hline
$E_{\rm T}^{\rm miss} < 50$~GeV &Above-Z &0&0&3& $7.3 \pm 2.0$ \\
&Below-Z &0& 0& 26& $25.0\pm 6.8$ \\
\hline
\rowcolor{LightGray}
$H_{\rm T} < 200$~GeV&& &&  & \\
\hline
$E_{\rm T}^{\rm miss} > 100$~GeV &Above-Z &2.01&0.92& 18& $13.0\pm 3.5$ \\
&Below-Z &0.13 &0& 21& $24\pm 9$ \\
\hline
$50<E_{\rm T}^{\rm miss} < 100$~GeV &Above-Z &0.14&0& 50& $46.0 \pm 9.7$ \\
&Below-Z &0&0& 142& $130 \pm 27$ \\
\hline
$E_{\rm T}^{\rm miss} < 50$~GeV &Above-Z &0.16&0&178& $200\pm 35$ \\
&Below-Z &0&0&  510& $560 \pm 87$ \\
\hline
\hline
\end{tabular}
 \end{center}
\end{table}

\begin{table}[t!]
 \begin{center}
\caption{Simulated number of events in our model for the 8~TeV run with
  an integrated luminosity of 19.5~fb$^{-1}$. Here, we set $m_{W_R} =
  2$~TeV, $g_R =0.4$, ${\cal R}_{e1} = 10^{-5}$,
  and $\tan \beta =1$. }
\label{tab:simulationmi5}
\vspace{5pt}
\begin{tabular}{ll|cccc}
\hline
\hline
Category & $m_{\ell^+ \ell^-}$& $m_{N_1} = 1$~TeV & 1.6~TeV &
Observed& Expected \\ \hline
\rowcolor{LightGray}
$H_{\rm T} > 200$~GeV
&& &  && \\
\hline
$E_{\rm T}^{\rm miss} > 100$~GeV &Above-Z &4.76 & 1.67&5 &$3.6\pm 1.2$ \\
&Below-Z &0 & 0& 7& $9.7\pm 3.3$ \\
\hline
$50<E_{\rm T}^{\rm miss} < 100$~GeV & Above-Z &0.60 &0.03& 4& $5.0 \pm 1.6$ \\
&Below-Z &0&0& 10& $11.0\pm 3.8$ \\
\hline
$E_{\rm T}^{\rm miss} < 50$~GeV &Above-Z &0&0&3& $7.3 \pm 2.0$ \\
&Below-Z &0& 0& 26& $25.0\pm 6.8$ \\
\hline
\rowcolor{LightGray}
$H_{\rm T} < 200$~GeV&& &&  & \\
\hline
$E_{\rm T}^{\rm miss} > 100$~GeV &Above-Z &5.53&1.81& 18& $13.0\pm 3.5$ \\
&Below-Z &0.38 &0& 21& $24\pm 9$ \\
\hline
$50<E_{\rm T}^{\rm miss} < 100$~GeV &Above-Z &0.44&0& 50& $46.0 \pm 9.7$ \\
&Below-Z &0&0& 142& $130 \pm 27$ \\
\hline
$E_{\rm T}^{\rm miss} < 50$~GeV &Above-Z &0.47&0&178& $200\pm 35$ \\
&Below-Z &0&0&  510& $560 \pm 87$ \\
\hline
\hline
\end{tabular}
 \end{center}
\end{table}

In Table~\ref{tab:simulation} and \ref{tab:simulationmi5}, we show the
number of events in each category simulated in our analysis for the
8~TeV run with an integrated luminosity of 19.5~fb$^{-1}$. Here, we set
$m_{W_R} = 2$~TeV, $g_R =0.4$, $\tan \beta =1$ and ${\cal R}_{e1} =
10^{-3}$ (${\cal R}_{e1} = 10^{-5}$) in Table~\ref{tab:simulation}
(Table~\ref{tab:simulationmi5}). We show the results for
two cases, $m_{N_1} = 1$ and 1.6~TeV. It turns out that our model
prediction is consistent with the current data. Moreover, we find that
our model potentially accounts for a small deviation from the SM
prediction in the $H_{\rm T} < 200$~GeV, $E_{\rm T}^{\rm miss} >
100$~GeV, and $m_{\ell^+ \ell^-} > 105$~GeV category without conflicting
with the results in the other categories. This observation indicates
that the trilepton plus missing energy search at the LHC Run-II will
offer a promising way to test our scenario in the near future, together with
other $W_R$ searches.

\section{Conclusion and Discussions}
\label{sec:conclusion}

In this paper, we have discussed an extended gauge sector model based on
the $\text{SU}(2)_L \otimes \text{SU}(2)_R \otimes \text{U}(1)_{B-L}$
gauge theory which accommodates the inverse seesaw structure in the
neutrino sector. We have found that our model can explain the CMS $eejj$
anomaly and the ATLAS diboson excess simultaneously, without conflicting
with existing experimental bounds. To explain these two anomalies, we
need sizable left-right mixing in the gauge sector. Such left-right
mixing can also appear in the neutrino sector because of the inverse
seesaw structure. This
allows us to probe our model in the searches for the  trilepton plus
missing energy signatures. After all, we expect that the LHC Run-II experiments
will test our setup in the near future and shed light on the nature of
TeV-scale physics beyond the SM.

\section*{Acknowledgments}

The work of N.N. was supported by DOE grant No.~DE-SC0011842 at the
University of Minnesota. The work of N.O. was supported by DOE grant
No.~DE-SC0013680. 




\end{document}